\documentclass[amsmath,twocolumn,aps,prl,superscriptaddress,floatfix]{revtex4-1}

\usepackage{amsmath}
\usepackage{amsfonts}
\usepackage{graphicx}
\newcommand{\beq}{\begin{equation}}
\newcommand{\eeq}{\end{equation}}

\newcommand{\beqar}{\begin{eqnarray}}
\newcommand{\eeqar}{\end{eqnarray}}

\newcommand{\J}{\mathcal J}

\newcommand{\Eq}[1]{Eq.~(\ref{#1})}
\newcommand{\Eqs}[1]{Eqs.~(\ref{#1})}
\newcommand{\nuom}[1]{\nu(\omega_{#1})}

\newcommand{\la}{\langle}
\newcommand{\ra}{\rangle}

\begin{document}

\title{Non-Makovian decoherence of a two-level system weakly coupled to a bosonic
bath}

\author{A.A. Slutskin }

\author{ K.N. Bratus'}
\affiliation{B.Verkin Institute for Low Temperature Physics and Engineering, Kharkov,
Ukraine}
\author{A. Bergvall}
\affiliation{Chalmers University of Technology, S-412 96 G\"oteborg, Sweden}
\author{V.S. Shumeiko}
\affiliation{Chalmers University of Technology, S-412 96 G\"oteborg, Sweden}
\date{21 May 2011}

\begin{abstract}
Bloch-Redfield equation is a  common tool for studying evolution of qubit systems weakly coupled to environment.  We investigate the accuracy of the Born approximation underlying this equation. We find that the high order terms in the perturbative expansion contain accumulating divergences that make straightforward Born approximation inappropriate. We develop diagrammatic technique to formulate, and solve the improved self-consistent  Born approximation. This more accurate treatment reveals an exponential time dependent prefactor in the non-Markovian contribution dominating the qubit long-time relaxation found in Phys. Rev. B {\bf 71},  035318 (2005). At the same time, the associated dephasing is not affected and is described by the Born-Markov approximation. 
\end{abstract}

\maketitle

%
{\bf Introduction.}
A quantum two-level system coupled to a bath of harmonic oscillators is a prominent model widely employed  for describing dissipation in quantum physics. The spin-boson model is outlined in several textbooks \cite{Weiss, Petruccione, Blum}, the advances are covered  by the review articles  (e.g. \cite{Caldeira, Leggett, Skinner} and references therein). Immense literature is devoted to applications of the model to  virtually all physics areas ranging from atomic physics and quantum optics to chemical physics and solid state physics.  With the advent of quantum information the model regained increased attention, particularly within the field of  quantum superconducting  circuits \cite{Makhlin}.   

For the qubit applications, the most interesting is the weak coupling regime when the 
two-level system slowly looses coherence and relaxes to the equilibrium state. The 
Bloch-Redfield equation \cite{Bloch,Redfield} is a common tool for describing this 
process. It is valid in the lowest, second order approximation with respect to the 
coupling to the bath, and employs the Markov approximation \cite{Blum}.  Several 
schemes beyond these approximations have been worked out based on the 
projection \cite{Petruccione}, path integral \cite{Leggett,Weiss}, diagrammatic \cite
{Makhlin2} and renormalization group \cite{Schoeller} techniques. Lifting constraints of 
the spin-boson model,  such as a linear coupling to the bath \cite{Makhlin3}, 
or equilibrium state of the bath \cite{Galperin,Hu}, 
have been discussed in literature.  However, in the 
qubit research, the Bloch-Redfield equation remains the basic theoretical model, 
whose prediction about an exponential in time qubit evolution is considered to be 
qualitatively correct.  To what extent is this true? The negative answer was obtained 
in \cite{Loss}, where it was found that the qubit long-time  relaxation is governed, 
within the Born approximation,  by a power-law time dependent non-Markovian term. 
This phenomenon is of a fundamental origin being related to a bounded 
energy spectrum of the bath that generates singular branching points of the qubit Green 
function \cite{Khalfin}. 

In this paper we revisit the problem of long-time decoherence in the spin-boson 
model in the weak coupling limit. We analyze the perturbation expansion and find 
that the high-order corrections to the 
Born approximation contain accumulating divergences, which make straightforward 
Born approximation inappropriate and require summation of the whole perturbation 
series.
To identify dangerous perturbative terms and perform the summation, we develop a relatively simple diagrammatic technique for the Liouville superoperator, and formulate an improved, self-consistent Born approximation. Within this approximation we find that the Green function branching points are shifted from the real axis yielding an exponential time dependent prefactor in the non-Markovian decoherence. For the relaxation, the corresponding rate is smaller than the Born rate, $1/T_1$, and thus the non-Markovian term dominates the long-time relaxation. For the related dephasing, on the contrary, the Born rate, $1/2T_1$, is larger, and the result of the Bloch-Redfield equation remains unchanged.  \\


{\bf Perturbative expansion.}
 The Hamiltonian of the spin-boson model has the form
\begin{equation}
 \hat H =  H_a + H_b + H_{ab},
\end{equation}
 where $ H_a = (\Omega/2)\, \sigma_z$ is the Hamiltonian of the two-level system,
 $\quad H_b =\sum_k\omega_k b^+_k b_k$%
is the Hamiltonian of the bosonic bath, and
 $H_{ab}= \hat\gamma \sum_k \eta_k\, (b^+_k + b_k)$
 is the coupling Hamiltonian,
$\hat\gamma=\gamma_z\sigma_z + \gamma_x\sigma_x \ll \hat 1$ is the coupling coefficient matrix containing the
longitudinal and transverse components. We note that in what follows we do not use the rotating wave approximation.


We describe the evolution of the full density matrix, $\rho(t)$, with the Liouville 
equation in the Laplace form,
\begin{equation}\label{rho_lambda}
\lambda  \rho(\lambda) + \check{\cal L} \rho(\lambda)  = -i\rho(0), \quad
\rho(\lambda)=\int_0^\infty dt\,e^{-i\lambda t} \rho(t),
\end{equation}
where $\check{\cal L}$ is the Liouville (super)operator defined through  $\check{\cal L}\,\ldots = [H,\,\ldots]$,
$\rho(0)$ is the initial condition, which we assume in the factorized form, $\rho(0)=
\rho_a(0) \rho_b$, with the equilibrium density matrix of the bath $ \rho_b =
Ze^{-H_b/T}$. The master equation for the  reduced density matrix of the two-level
system, $\rho_a(\lambda) =  \mathrm{Tr}_{b}\rho(\lambda)$, is known to  have the
form \cite{Petruccione},
\begin{equation}\label{rho_a_lambda}
\lambda  \rho_a(\lambda) + \check{\cal L}_a \rho_a(\lambda) -
 \check \J(\lambda)  \rho_a(\lambda)= -i\rho_a(0)\,,
\end{equation}
where the self-energy superoperator $\check\J(\lambda)$ represents the effect of the bath.
Our goal will be to derive a diagrammatic representation for the self-energy analogous to the one of the many-body diagrammatic
theory \cite{Abrikosov}. 

To this end we introduce the system+bath Green function, $\check g(\lambda)$
through the relation, $\rho(\lambda) = -i\check g(\lambda) \rho(0)$, and the Green function, $\check G(\lambda)$, of
the two-level system, $\rho_a(\lambda) = -i\check G(\lambda)\rho_a(0)$, implying 
$\check G(\lambda) = \la \check   g(\lambda)\ra \equiv  {\rm Tr}_b [\check g(\lambda)\rho_b]$ These Green functions are superoperators that satisfy the
equations,
\begin{eqnarray}\label{g}
\check g =\check g_0 - \check g_0 \check {\cal L}_{ab}\check g\,, \quad  
\check g_0(\lambda) = (\lambda + \check{\cal L}_a + \check{\cal L}_b )^{-1}, 
\end{eqnarray}
\begin{equation}\label{gG}
\check G = \check G_0 + \check G_0 \check \J \check G\,, \quad \! \! \check G_0(\lambda) = (\lambda + \check{\cal L}_a )^{-1}.
\end{equation}
Our diagrammatic approach is derived from the observation that the n-th order term of the perturbative expansion of the  Green function in \Eq{g},  $\check g = \check g_0 - \check g_0 \check {\cal L}_{ab}\check g_0 + \ldots$, can be exactly presented on the form,
 \begin{eqnarray}\label{rho_n1}
\check g^{(n)}(\lambda)\!
= \!\!\sum_{\{k,s\}}
 \prod_{r=1}^{n}\limits \eta_{k_r}\!
\check   G_{0}\!\!\left(\lambda \! + \!  \sum_{i=1}^{r}s_i
\omega_{k_i}\!\!\right)\! \check R_{k_r, s_r}  \check G_{0}(\lambda).
 \end{eqnarray}
Here  we introduced notations, $\check R_{k,s}\, \ldots = -\left[\hat\gamma \,b_{ks} ,\, \ldots\right]$, and  
$b_{k+}= b^+_k , \, b_{k-}=b_k$; the product is ordered, and the summation runs over all $k$ and $s=\pm$. To justify this equation we consider the action of each term of the expansion on some operator of the system, $X_a$. The action of the first term reads,  $\check g_0(\lambda) X_a =\check G_0(\lambda) X_a$.  The second term follows from the identity,  $\check g_0(\lambda)b_{ks}\check g_0(\lambda) X_a =
\check G_0(\lambda+s\omega_k) b_{ks} \check G_0(\lambda) X_a$. Continuing this procedure we arrive at \Eq{rho_n1}, and then obtain expansion for $\check G$ by performing the averaging, 
\begin{equation}\label{average}
\check G(\lambda) = \sum_n \la g^{(n)}(\lambda)\ra
\end{equation}
%

%
\begin{figure}
\includegraphics[width=0.42\textwidth]{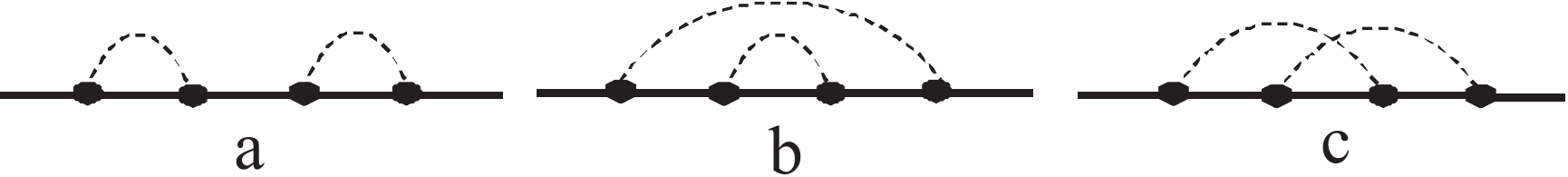}
\caption{Forth order diagrams for the Green function $\check G$: lines represent $G_0$, generally
with shifted arguments; dots represent vertices $R$;  dashed lines depict parings.}\label{G4}
\end{figure}

Averaging in \Eqs{average}, (\ref{rho_n1}) is illustrated graphically in Fig.~\ref{G4}: The lines represent unperturbed Green functions, $\check G_0$, generally with shifted arguments, the dots represent superoperatores $\check R$, and the dashed lines represent the parings of boson operators.  All the diagrams split into  the two classes: reducible, similar to the one in Fig.~\ref{G4}a, and irreducible,
Fig.~\ref{G4}b,c. Since the parings select coinciding vertex indices, $k_{r'} = k_r$, and $s_{r'} =
-s_r$, any line of the reducible diagram that connects irreducible segments corresponds
to $\check G_0(\lambda)$ with {\em non-shifted} argument. This property also refers  to the edge
lines. Thus  any n-th order diagram has the form, $ \check G_{0}(\lambda)\la
\check S_{2m_1}\ra \check G_{0}(\lambda) \ldots \check G_{0}(\lambda) \la \check S_{2m_r}\ra \check G_{0}(\lambda)$, where $\langle \check S_{2m_i}\ra$ are  irreducible segments,
$2m_1 + \cdots + 2m_r = n$.%

\begin{center}
\begin{figure}
\includegraphics[width=0.45\textwidth]{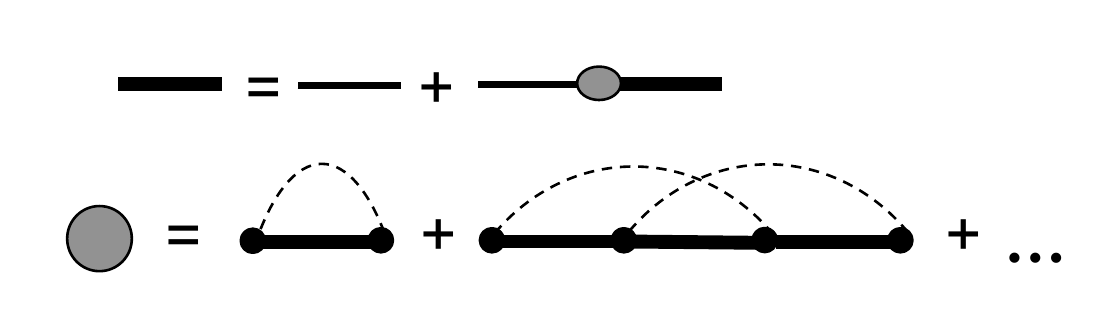}
\caption{Dyson equation and self-consistent self-energy diagrams: bold lines indicate
full Green functions $G(\lambda)$, bold circle indicates self-energy $\J(\lambda)$
containing only crossing diagrams.}\label{Dyson}
\end{figure}
\end{center}

The latter property allows us to formulate an efficient diagrammatic theory and present equation for the Green function 
$\check G(\lambda)$ on the form of the Dyson equation, similar to conventional diagrammatics \cite{Abrikosov},
\begin{equation}\label{GDyson}
\check G(\lambda) = \check G_0(\lambda)+ \check G_0(\lambda)\la \check S(\lambda) \ra \check G(\lambda),
\quad \la \check S(\lambda)\ra  = \sum_{m=1}^\infty \la \check S_{2m}\ra\,,
\end{equation}
here  the summation goes over all irreducible diagrams. Comparison of \Eqs{gG} and (\ref{GDyson}) establishes explicit form for the self-energy, $\check \J(\lambda) = \la \check S(\lambda)\ra$. The Born approximation corresponds to the first term ($\sim \gamma^2$) in expansion (\ref{GDyson}),
\begin{eqnarray}\label{JBorn}
\check {\cal J}_B(\lambda) = \la \check S_2\ra = \sum_{k,s} \eta^2_k\langle \check R_{k,-s}\check G_0(\lambda +
s\omega_k)\check R_{k,s}\rangle\,.
\end{eqnarray}
One may make one step further, and formulate an improved perturbative expansion by regrouping  the 
series terms as shown in Fig.~\ref{Dyson} \cite{Abrikosov}. Here the bold lines (full Green functions $\check G$) replace the thin lines (free Green functions $\check G_0$)  in all the self-energy diagrams, and only the diagrams with crossing pairing lines remain. In fact, such a rearrangement is necessary because of the presence of accumulating divergences in the non-crossing diagrams similar to the ones in Fig.~\ref{G4}b (see below).

%

The Dyson equation (\ref{gG}) with the self-consistent diagrammatic representation for
the self-energy in Fig.~\ref{Dyson} is the main technical result of this paper. The lowest order, 
self-consistent Born approximation is now represented with the first diagram in Fig.~\ref{Dyson} 
and is given by \Eq{JBorn} with the
full Green function replacing the free Green function, $\check G_0 \rightarrow \check G$.

For practical calculations, the vector form of the density matrix is convenient, then the
self-energy becomes the $4\times 4$ matrix. Using the Pauli matrix bases,
$(1,\sigma_z,\sigma_+, \sigma_-)$, $ \sigma_\pm = (\sigma_x \pm i\sigma_y)/2$, in which
the Liouvillean of the system acquires the diagonal form, $\check {\cal L}_a \rightarrow
\mathrm{diag} (0,\,0,\,\Omega,\,-\Omega\,)$, the self-energy matrix takes the form,
\begin{eqnarray}\label{JBorn2}
{\cal J}_B(\lambda) &= & \sum_{s}\int_0^\infty d\omega \,\nuom{} P_+
G(\lambda + s\omega_k) 
\times \nonumber\\ &&
\left[ P_+ \coth(\omega / 2T) + sP_-\right]\,.
\end{eqnarray}
Here we introduced the bath spectral density, $\nu(\omega) = \sum_k\eta^2_k\delta(\omega
- \omega_k)$; the matrices $P_\pm$ read,
\begin{equation}\label{P}
P_+ = \left(
\begin{array}{cc}
0 & \gamma_x p_+ \\
2\gamma_x \overline p_+ & \gamma_z\sigma_z \\
\end{array}%
\right), %
\;\; %
P_- = \left(
\begin{array}{cc}
\gamma_z\sigma_x  & \gamma_x p_- \\
2\gamma_x \overline p_- & 0\\
\end{array}%
\right), %
\end{equation}
where $p_\pm =(1/4) [1  \mp (\sigma_z + 2\sigma_{\mp})]$, and bar 
indicates conjugated matrix.\\

{\bf Transverse coupling.} 
We first analyze \Eqs{JBorn2}, (\ref{P}) for the purely transverse coupling assuming 
$\gamma_z=0$.
In this case 
the $P$-matrices in \Eq{P} acquire a block-antidiagonal form, which induces the block-diagonal form of solution to the 
self-consistent equations (\ref{gG}) and (\ref{JBorn2}), namely, $G = \mathrm{diag}(G_\parallel,\, G_\perp)$, and 
$\J_B = \mathrm{diag}(\J_\parallel,\, \J_\perp)$.  Correspondingly, the master equation for the density matrix splits into the two independent parts:
equation for $\rho_z$ governed by $\J_\parallel$ (relaxation), and equation for
$\rho_\pm$ governed by $\J_\perp$ (dephasing). Furthermore, the functional dependence 
between $\J_B$ and $G$ acquires peculiar structure,
\begin{equation}\label{ParPerp}
\J_\parallel(\lambda)=\J_\parallel[G_\perp(\lambda)], \quad
\J_\perp(\lambda)=\J_\perp[G_\parallel(\lambda)],
\end{equation}
which allows us to fully  investigate the analytical properties of the Green functions and thus 
describe the long-time evolution of the density matrix. 

\begin{center}
\begin{figure}[h!]
\includegraphics[width=0.45\textwidth]{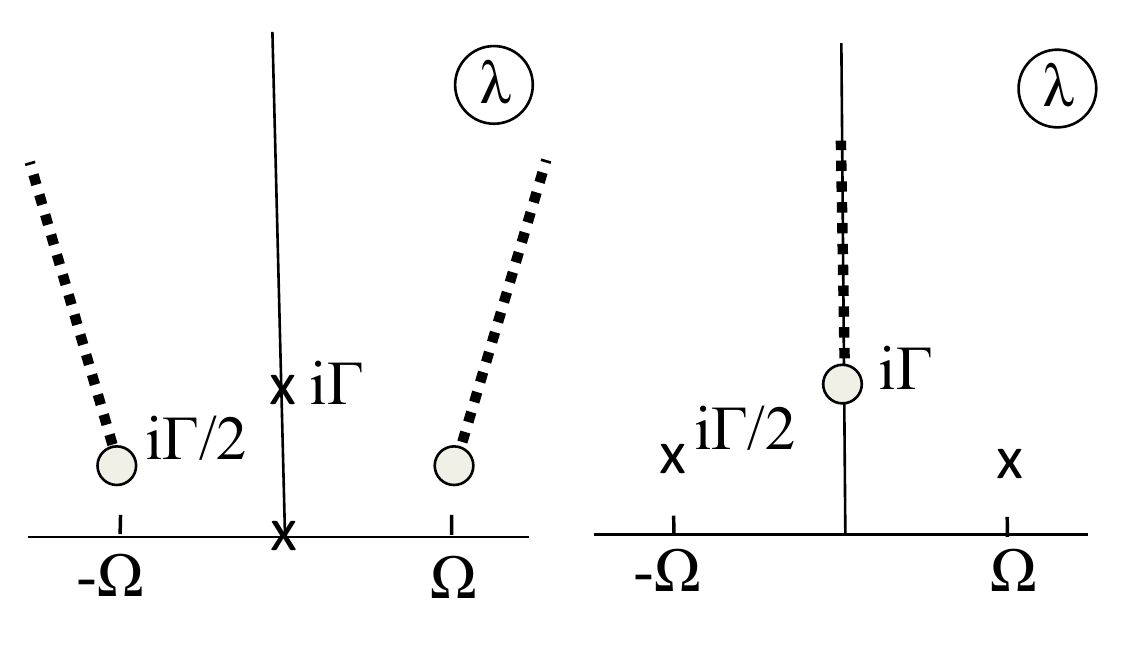}
\caption{Singular points of the longitudinal (left) and transverse (right) Green
functions: poles are indicated with crosses, branching points with circles, branch cuts
are shown with dashed lines. }\label{Contours}
\end{figure}
\end{center}

The analytical properties of the Green functions, $G(\lambda) = (\lambda + {\cal L}_a - \J_B(\lambda))^{-1}$, are determined by the singular points, which consist of the poles and the branching points stemming from the self-energy singularities. Let us first discuss the positions of the poles; they are given by equation, Det$(\lambda +
{\cal L}_a - \J_B(\lambda))=0$. For the longitudinal Green function $G_\parallel$, the
determinant is degenerate, and the equation has two roots, $\lambda^\ast =0$, and
$\lambda^\ast=J_\parallel(\lambda^\ast)$, where $J_\parallel=(\J_\parallel)_{22}$ is the
diagonal element of the self-energy. Since $\J_B(\lambda)\sim \gamma_x^2$ is small,
the second root is small provided $J_\parallel$ is regular at $\lambda=0$. Then the
approximate position of this pole is $\lambda_\parallel^\ast = J_\parallel(0)$, see Fig.
\ref{Contours}. Similar consideration applies to the transverse Green function
$G_\perp$, whose poles in the main approximation are $\lambda_\perp^\ast = \mp\,\Omega
+J_\perp(\pm\,\Omega)$ ($J_\perp$ is the diagonal element of $\J_\perp$).
Straightforward evaluation of the pole positions using approximation, $G\approx G_0$, in
\Eq{JBorn2}, gives,
\begin{equation}\label{}
\lambda_\parallel^\ast=  2\pi i\gamma_x^2 \nu(\Omega)\coth{\Omega\over 2T} \,\equiv\,
i\Gamma, \quad \lambda_\perp^\ast= \pm\,\Omega + {i\Gamma\over 2}
\end{equation}
(in the latter equation, small correction to the real part was omitted).
%
%
%

The branching points appear in the self-energy as the result of integration over
$\omega$ in \Eq{JBorn2}  of the singular Green function: Since the integral is defined
on the semi-axis, $\omega\geq 0$, the singularity builds up when the Green function pole approaches the
integration edge $\omega=0$. Consequently, the branching points,
$\lambda_\parallel^\circ$, of $\J_\parallel$ coincide, by virtue of 
\Eq{ParPerp}, with the poles of the transverse Green functions,
$\lambda_\parallel^\circ = \lambda_\perp^\ast$, and vice versa for the transverse
part, $\lambda_\perp^\circ = \lambda_\parallel^\ast$ as shown on Fig.
\ref{Contours}. This verifies  the above made assumption regarding analytical behavior of the
self-energy near the poles of corresponding Green function.

Important conclusion of our analysis is that both the poles and branching points lie in
the complex $\lambda$-plane. Therefore the time evolution of the density matrix is always {\em
exponential}. This conclusion is the main result of the paper. 

Within the straightforward Born approximation of \Eq{JBorn}, the branching points lie on the real
axis being defined by the poles of non-perturbed Green functions, $G_0$. This leads to 
the non-exponential time relaxation found in \cite{Loss}. However, the higher order diagrams also contain the singularities that accumulate
with the growing diagram order. The most dangerous diagrams are of the type 
of Fig.~\ref{G4}b, having the singularities, $\gamma_x^{2m}/(\lambda-\lambda^\ast)^{m}$.  This shortcoming of the straightforward perturbation theory is eliminated by the improved perturbative expansion.


A more accurate treatment within the self-consistent Born approximation reveals the shifts of the
self-energy branching points into the complex plane.  In the case of $G_\parallel$, they situate more close to the real axis compared to the pole, left panel in Fig.~\ref{Contours}, and  thus define the the long-time relaxation. Explicit solution for $\rho_z(t) $ has the form,
\begin{eqnarray}\label{rhozt}
\rho_z(t) =  \!\int_{-\infty-i0}^{\infty-i0}
\frac{d\lambda}{2\pi i}\left[\frac{e^{i\lambda t} \rho_z(0)}{\lambda - J_\parallel}
+\frac{e^{i\lambda t}(\J_\parallel)_{21}}{2\lambda(\lambda - J_\parallel)}\right].
\end{eqnarray}
The contribution of the poles reads,
\begin{equation}\label{rhozp}
\rho_z^{(p)}(t) = \rho_z(0)e^{-\Gamma t}+ {1\over 2}\tanh \frac{\Omega}{2T} (e^{-\Gamma
t} - 1).
\end{equation}
This coincides with the result of the Markovian approximation of the Bloch-Redfield theory
\cite{Redfield}, $\Gamma= 1/T_1$. The result of integration along the branch cuts depends on the specific 
form of the bath spectral density. Here we assume the ohmic bath for certainty,  
$\nu(\omega) = K\omega e^{-\omega/\omega_c}$, with a large cutoff frequency,  $\omega_c \gg
\Omega \gg T$.  The asymptotical contribution of the branch cuts  at $\Omega t\gg 1$ reads,
\begin{equation}\label{rhozc}
\rho^{(c)}_z(t) = -  \, 4\gamma_x^2 K\,{\cos\Omega t\over (\Omega t)^2}\, e^{-(\Gamma/2)
t}F(T,t)\,\rho_z(0),
\end{equation}
where
\begin{eqnarray}\label{}
F(T,t) &=& {(\pi Tt)^2\over\sinh^2(\pi Tt)} + {2T\over \omega_c} f(2Tt),\\
f(x) &=& x^2\int_0^{\pi/2}dy\,e^{-xy} y^2\cot y.\nonumber
\end{eqnarray}
The function $F(T,t)$ approaches the asymptotic values, $F=1$ in the limit $ Tt\ll
1$, and $F=2T/\omega_c $ at $Tt\gg 1$. 


Although the contribution from branch cuts in \Eq{rhozc} decays slower than the Markovian part, \Eq{rhozp}, it contains a small rapidly oscillating factor, and because of this, the non-Markovian part becomes dominant at rather large times,
\begin{equation}
t \geq 4T_1 \ln {\Omega t\over \gamma_x} > T_1\,.
\end{equation}
The oscillation, $\propto \cos\Omega t$, formally results from the positions of the branching points, defined by the poles of $G_\perp$, away from the imaginary axis. It can therefore be interpreted as the effect of coupling to the transverse evolution (cf. \Eq{rho+} below).

The situation is different for the dephasing. In this case the branching point of $G_\perp$, right panel in Fig.~\ref{Contours}, has larger imaginary part than the poles, and thus non-Markovian contribution decays more rapidly.  Therefore the long-time dephasing in the case of transverse coupling is {\em  given by the Markov approximation},
\begin{eqnarray}\label{rho+}
\rho_+^{(p)}(t) &=& \rho_+(0) e^{
-i\Omega t
 -(\Gamma/2) t},\nonumber\\
\rho_+^{(c)}(t) &=&  -\frac{2\gamma^2 K}{(\Omega t)^2}e^{-\Gamma
t}\left(\rho_+(0)  + \rho_-(0)\right).
\end{eqnarray}
%

{\bf  General coupling.}
The presence of both the transverse and longitudinal couplings, $\gamma_x,\, \gamma_z\neq 0$, in \Eq{P} leads to the two complications: First, the block-diagonal form of the Green functions is lost due to the fact that the $P$-matrices acquire diagonal parts,  and the relaxation and dephasing become mixed. Second, the property, \Eq{ParPerp}, is lost, which implies more complex structure of the 
singularities consisting of the coinciding poles and branching points. Attentive 
analysis, however,  shows that the mixing of the relaxation and dephasing is of the higher order, 
$\sim\gamma^4$,
and within the Born approximation, this effect can be neglected. Furthermore, 
the property, \Eq{ParPerp}, is only violated for the transverse (dephasing) part of the self-energy,
\begin{equation}\label{Jperp}
\J_\perp = \J_\perp[G_\parallel] +  \J'_\perp[G_\perp],
\end{equation}
while for the longitudinal part it persists. Thus, up to the fourth-order corrections, the relaxation is not affected by the presence of $z$-coupling.

For the dephasing, the major problem is the contribution of the singular points at 
$\lambda = \pm\Omega + i\Gamma/2$, affected by the second term in \Eq{Jperp}. To evaluate this contribution we note that the first term in \Eq{Jperp}, $\J_\perp[G_\parallel] \propto \gamma_x$, is related to the transverse coupling, and it is analytical at this point and can be evaluated as before, $\J_\perp[G_\parallel] \approx \J_\perp(\pm\Omega)= i\Gamma/2$. Thus the role of this term is to shift the frequency, $\sigma_z\Omega\rightarrow \sigma_z\Omega - i\Gamma/2$, in the equation for the transverse Green function,
\begin{equation}\label{Gxz}
G_\perp(\lambda) = \left( \lambda + \sigma_z\Omega - i\Gamma/ 2 -
\J'_\perp(\lambda)\right)^{-1}.
\end{equation}
This equation, however, apart from the frequency shift, describes the effect of pure $z-$coupling (since $\J'_\perp \propto\gamma_z$), and
the solution to this case is well known in the literature (e.g. \cite{Makhlin} and references therein). Therefore the dephasing in the present case is given by this solution with the shifted frequency,
\begin{eqnarray}\label{rhopure}
\rho _+(t) \!\!\! &=& \!\!\! e^{ -i\Omega t -(\Gamma/2)t} F(t)\,\rho_+(0)\, ,\nonumber\\
F(t) \!\!\! &=& \!\!\! \exp\left(- \gamma_z^2\int^{}_{} d\omega\,{\nu(\omega) \over\omega^2}
\sin^2\frac{\omega t}{2} \coth \frac{\beta\omega }{2}\right).
\end{eqnarray}
%

{\bf Conclusion.} We investigated long-time decoherence of a two-level system weakly coupled to a bosonic bath. We found that the conventional Born approximation does not correctly describe the evolution of the system, and formulated an improved, self-consistent  Born approximation based on a developed diagrammatic technique. We found an exponential time dependent prefactor in slow varying non-Markovian tail that dominates the long-time relaxation, while the dephasing associated with  relaxation is well described within the Markov approximation. \\

{ \bf Acknowledgement.} Support from FP-7  MIDAS Consortium, and Swedish Royal Academy (KVA) is gratefully acknowledged.


\end{document}